\documentstyle[floats,twocolumn,aps,prl]{revtex}

\begin{document}
\ifpreprintsty\else
\twocolumn[\hsize\textwidth%
\columnwidth\hsize\csname@twocolumnfalse\endcsname
\fi
\title{Influence of disorder and a parallel magnetic field
on a Quantum Cascade Laser}
\author{V.M. Apalkov and Tapash Chakraborty}
\address{Max-Planck-Institut f\"ur Physik Komplexer Systeme, 
01187 Dresden, Germany}
\maketitle
\begin{abstract}
The luminescence spectra of a quantum cascade laser in a strong
magnetic field is influenced significantly by the presence of
disorder (charged or neutral) in the system. An externally applied
magnetic field parallel to the electron plane causes a red shift of
the luminescence peak in the absence of any disorder potential.
Our results indicate that the disorder potential tends to cancel
that red shift and causes a rapid decrease of the luminescence
peak. A similar behavior was observed in a recent experiment
on QCL in a parallel magnetic field.
\end{abstract}
\ifpreprintsty\clearpage\else\vskip1pc]\fi
\narrowtext
Since the invention of the unipolar semiconductor quantum cascade
laser (QCL) in 1994 \cite{first,vertical}, there have been a
series of innovations and improvements of performances in this
field of intersubband light sources 
\cite{shortwave,room,aboveroom,gaas,strasser,proceed}.
The QCL has already established its potential as a tunable,
high-power coherent source in the mid-IR spectral range, and
therefore amenable to a host of very useful applications \cite{proceed}.
In the quest for further improvements of the QCL device, magnetic
field studies have recently been performed \cite{blaser1,blaser2}.
However, no experimental or theoretical studies are available yet
on the role of disorder in the optical properties of the QCL.
Although the techniques of crystal growth have improved significantly
in recent decades, interface roughness in QCL structures cannot be
entirely eliminated. In a narrow quantum well, electrons are closer
to the interface and are influenced by the irregularities or by
the charged impurities confined at the interface. In this letter,
we report on our findings that the effect of disorder is particularly
significant on the luminescence spectra in the presence of a magnetic 
field. The QCL structure we have studied here is sketched in 
Fig. 1 \cite{qcltilt}, where the parameters are taken from the
QCL structure reported in \cite{blaser1}.

\begin{figure}
\begin{center}
\begin{picture}(120,130)
\put(0,0){\includegraphics{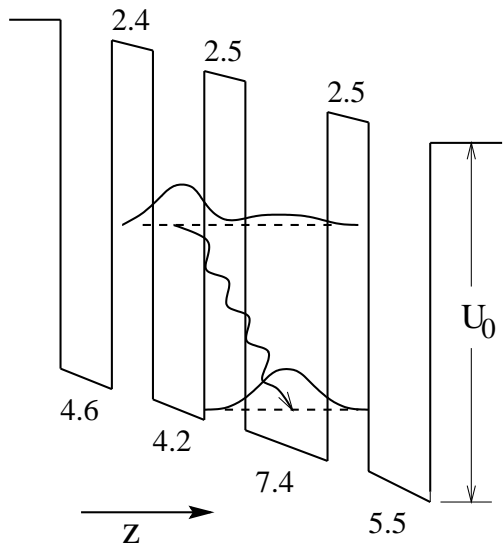}}
\end{picture}
\caption{Energy band diagram (schematic) of the active region of
a quantum cascade laser structure under an average applied electric
field of 55 kV/cm. Only one period of the device is shown here.
The relevant wave functions (moduli squared) as well as
the transition corresponding to the laser action are also shown
schematically. The numbers (in nm) are the well (Ga$_{0.47}$In$_{0.53}$As)
and barrier (Al$_{0.48}$In$_{0.52}$As) widths.
Material parameters considered in this work are: electron
effective mass $m_e^*$ (Ga$_{0.47}$In$_{0.53}$As)=0.043 $m_0$,
$m_e^*$ (Al$_{0.48}$In$_{0.52}$As)=0.078 $m_0$, the conduction band
discontinuity, $U_0=520$ meV, the nonparabolicity coefficient,
$\gamma_w=1.3\times10^{-18}$ m$^2$ for the well and
$\gamma_b=0.39\times10^{-18}$ m$^2$ for the barrier, and the sheet
carrier density induced by doping, $n_s=2.3\times10^{11}$
cm$^{-2}$. The energy difference between the two levels where the
optical transition takes place, is 132 meV. 
}
\label{device}
\end{center}
\end{figure}

The main results of the experimental work on the optical spectra 
of a QCL in a strong parallel magnetic field \cite{blaser1} were,
a small blue shift of the emission peak, decrease of 
the emission intensity and broadening of the peak in a strong magnetic 
field. It was also found that the single-particle theoretical model 
can not explain these observations \cite{blaser1,blaser2}. This
model predicts a red shift of the peak that can not be cancelled 
by the many-body corrections \cite{qcltilt}. In order to explain the 
experimentally observed tiny blue shift of the emission line,
we consider the single-particle Hamiltonian
\begin{eqnarray}
H & = &
\frac{p_z^2}{2m(E)} + \frac{1}{2m(E)} \left(
    p_x +\frac{e}{c}Bz \right) ^2 + \frac{p_y^2}{2m(E)}  \nonumber \\
     && + V_{\rm conf}(z) + V_{\rm dis}(x,y)
\label{hamil}
 \end{eqnarray}
where the magnetic field $B$ is applied in the $y$-direction, the 
vector potential is $\vec{A}=(Bz,0,0)$, $V_{\rm conf}(z)$ is 
confinement potential due to conduction-band discontinuity and bias 
electric field (Fig. 1), and $V_{\rm dis}(x,y)$ is the disorder 
potential. The effective mass of the particle depends on the energy due 
to the band nonparabolicity. We consider the same disorder potential 
for both quantum wells in the active region (Fig. 1), i.e., electrons 
feel the same potential before and after emission. 

We have studied the single-particle energy spectra and emission lines 
for two types of disorder potentials: for disorder due to the 
surface roughness in quantum wells and for disorder due to the 
doped charged impurities. The disorder potential in the first case is 
described by the potential $V_{\rm dis,1}(x,y)$ which has zero average 
and Gaussian correlations
\[
\left< V_{\rm dis,1}(\vec{r}_1)V_{\rm dis,1}(\vec{r}_2) \right> = 
  V_0^2 {\rm e}^{-(\vec{r}_1 -\vec{r}_2)^2/l_0^2}
\]
where $\vec{r}$ is the two-dimensional vector.  
  
In the case of a charged impurity we assume that all impurities are in 
the same two-dimensional plane with position $z_{\rm imp}$. The potential 
has the form
\[
V_{\rm dis,2}(\vec{r})= -\sum_{i}\frac{e^2}{\epsilon} 
 \frac{{\rm e}^{-|\vec{r}-\vec{R}_i|/d_{\rm sc}}}{\sqrt{
 (\vec{r}-\vec{R}_i)^2 + (z-z_{\rm imp})^2}}
\]
where $\vec{R}_i$ is the two-dimensional vector describing the position
of $i$-th impurity, $d_{\rm sc}$ is the phenomenological constant which 
describe the screening of impurity potential by electrons. In our
present calculations we used $d_{\rm sc}=10$ nm. The two-dimensional
density of impurities is equal to $n_{\rm imp} = 2.3\times 10^{11}$ 
cm$^{-2}$. The positions of impurities are uncorrelated.

To find the emission spectra of the system with the disorder
potential $V_{\rm dis}(x,y)$ we calculate from Hamiltonian
(\ref{hamil}) but without the disorder potential, the dispersion of 
the electron on the first and on the second subband (Fig. 1)
for a given magnetic field $B$. Without any disorder the states are 
characterized by the two-dimensional 
momentum, $(\hbar k_x, \hbar k_y)$. Taking these states as a basis we 
diagonalize the total Hamiltonian with the given disorder potential 
in a square geometry of size $L$ with periodic boundary conditions. 
The emission spectra are then found from 
\begin{eqnarray}
I(\omega) &=&\left\langle \sum_{if} \delta(\omega-E_i+E_f)
           \left|\int\,\xi_2(z)\xi_1(z) dz \right|^2\right. \nonumber \\
 && \left.\times \left| 
             \int \psi^{*}_i (k_x,k_y) \psi_f(k_x,k_y) dk_x dk_y \right|^2
              \right\rangle
\label{emit}
\end{eqnarray}
where $E_i$, $E_f$ are electron energies before and after emission;
$\xi_1(z)$, $\xi_2(z)$ are wave functions of the electron on the 
first and second subbands; $\psi_i(k_x,k_y)$, $\psi_f(k_x,k_y)$ are the
initial and final wave functions for a given disorder potential. The 
summation in (\ref{emit}) is over all final states and over occupied 
initial states. The number of occupied initial states is determined by 
$N_i = L^2/n_{\rm el}$, where $n_{\rm el} = 2.3\times 10^{11}$ cm$^{-2}$ 
is the electron density. The anglular brackets correspond to averaging 
over disorder potential.  

\begin{figure}
\begin{center}
\begin{picture}(120,130)
\put(0,0){\includegraphics{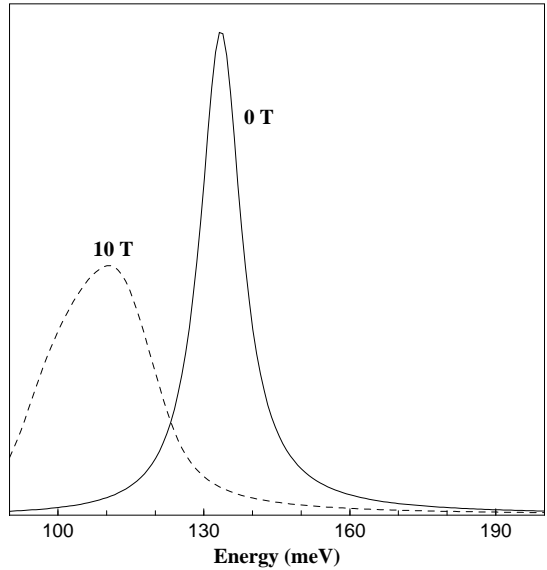}}
\end{picture}
\vspace*{4.0cm}
\caption{Luminescence spectra of a quantum cascade structure 
in the presence of an externally applied parallel magnetic field,
but in the absence of any impurity.}
\label{noimp}
\end{center}
\end{figure}

To understand the properties of the emission line of our system with 
disorder we consider the first terms of a perturbation theory in 
a magnetic field \cite{parallel}. In this case the Hamiltonian 
(\ref{hamil}) for subband $\alpha$ can be rewritten as
\[
H_\alpha = 
\frac{1}{2m(E)} \left(
    p_x +\frac{e}{c}B\langle z\rangle_\alpha \right)^2 + 
    \frac{p_y^2}{2m(E)} + V_{\rm dis}(x,y)
\]
where $\langle z\rangle_\alpha = \int \xi_\alpha(z)z\xi_\alpha(z) dz$ 
and we disregard the small 
diamagnetic term. In this case the energy spectra in the non-zero magnetic 
field is the same as for the zero magnetic field and the wave functions 
differ only by a phase factor which correspond to a translation in
$k$ space of the vector $\left(\frac{eB}{\hbar c}\langle z\rangle_\alpha,0
\right)$. The main effect is on the emission spectra because the 
intensity of optical transition from the initial state $i$ to the
final state $f$ is proportional to the overlap between 
$\psi_i(k_x,k_y)$ and $\psi_f(k_x-k_B,k_y)$, where $\psi_i$ 
and $\psi_f$ are the wave functions of the initial and final system 
respectively, in the absence of a magnetic field
\begin{equation}
{\cal I}_{if}=\left\vert\int\psi^*_i(k_x,k_y)\psi_f(k_x-k_B,k_y)
dk_xdk_y\right\vert^2.
\label{inten}
\end{equation}
The wave vector $k_B$ is equal to
$$k_B = \frac e{\hbar c}B(\langle z\rangle_i-\langle z\rangle_f)$$
and is of the order of $k_B\sim 0.1$ nm$^{-1}$ at $B=10$ tesla for
the system parameters considered here (Fig. 1). 

The position of the maximum of the emission line is determined mainly
by the transitions from the ground state of the single-particle
system. Without any disorder, the single-particle wave functions are 
plane waves $\psi_i(k_x,k_y)=\delta(k_x-k_{x,i})\delta(k_y-k_{y,i})$,
with the ground state at ${\vec k}=0$. Clearly, the allowed transitions 
are then only from the ground state ($k_x=0$) of the upper subband
to the excited states ($k_x=k_B$) of the lower subband, resulting
in a red shift of the emission line in a magnetic field. A strong 
enough disorder will localize the lowest states. If the localization 
length is smaller than $1/k_B$ then the overlap between the initial
ground state to the final ground state is large and transition 
from a ground state to a ground state (different subbands) of the system
is strong in a magnetic field and the red shift of the emission line
is suppressed.

\begin{figure}
\begin{center}
\begin{picture}(120,130)
\put(0,0){\includegraphics{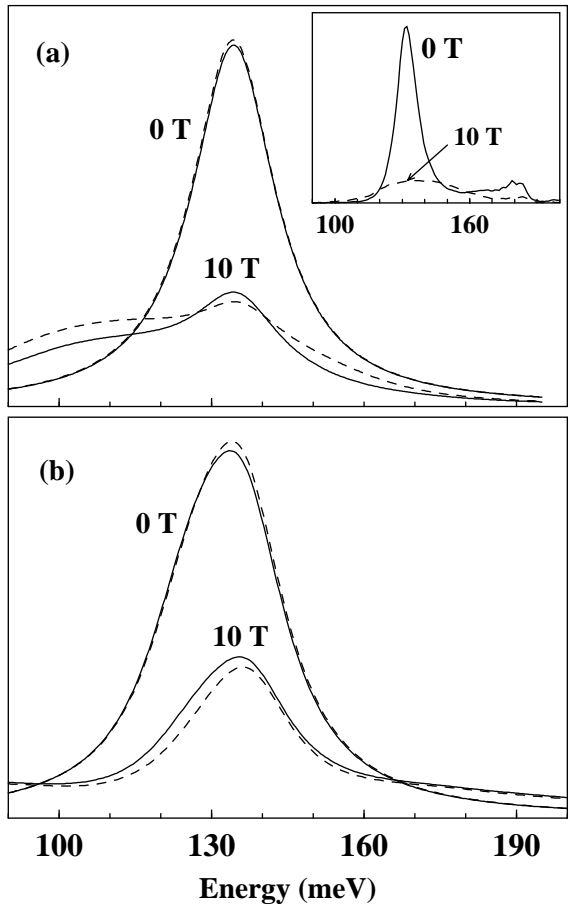}}
\end{picture}
\vspace*{8.0cm}
\caption{Luminescence spectra of a quantum cascade structure in
the case of (a) charge-neutral impurities and (b) charged impurities 
in the presence of an externally applied parallel magnetic field.
The experimental results of Ref.~\protect\cite{blaser1} are shown as
inset. }
\label{impure}
\end{center}
\end{figure}

In Fig. 2, the emission spectra are shown for the system without disorder
for two values of the magnetic fields. The large red shift, as expected,
is clearly seen in the figure. In Fig. 3, the emission spectra are shown 
for the system with disorder. In Fig. 3a, where the disorder is
charge neutral, the solid curves correspond to $V_0=30$ meV and $l_0=5$ 
nm, and the dashed curves correspond to $V_0=20$ meV and $l_0=5$ nm. 
In Fig. 3b, the disorder is due to the charged impurity and the solid 
and dashed lines correspond to separation between the impurity plane and 
$4.2$ nm quantum well (Fig. 1) to be 3 and 4 nm, respectively. 
In both cases we see large suppression of the red shift observed 
in Fig. 2. The emission line for the high magnetic field case
is almost at the same energy as for the zero-field case, and in 
Fig. 3b there is a small blue shift of the peak. These results are 
qualitatively in agreement with experimental data of 
\cite{blaser1,blaser2} and shown as inset in 
Fig. 3. Further, in Fig. 3 the intensity of the luminescence peak
shows a rapid
decrease in the presence of disorder and high fields because of
small overlap in Eq.\,(3).

The small blue shift in Fig. 3b is due to the band nonparabolicity. 
At zero magnetic field the nonparabolicity of band structure allows the 
transitions from the ground state to the exited states. The magnetic 
field redistributes the transitions and increase the intensity of 
transition into the ground state which results in small blue shift 
in Fig. 3b. From our results it is clear that in a QCL device, magnetic
field induced luminescence spectra \cite{qcltilt} can only be observed
in the absence of strong disorder, i.e., devices with high electron
mobility or for wider quantum wells.

We thank P. Fulde for his support and kind hospitality in Dresden
and St\'ephane Blaser for valuable discussions on his experimental
results.

\vfil
\end{document}